\documentclass[aps,nopacs,preprintnumbers,prb,twocolumn]{revtex4}
\usepackage{graphicx}
\usepackage{bm}
\usepackage{bbold}
\usepackage{amsmath}

\usepackage{epsfig}
\usepackage{physics}
\usepackage{epstopdf}
\usepackage{hyperref}

\usepackage{url}
\usepackage{color}
\usepackage{multirow}
\usepackage{comment}
\usepackage{mathrsfs}
\begin{document}
\title{A first-principles investigation of the origin of superconductivity in TlBi$_2$}
\author{Aiqin Yang}
\affiliation{MOE Key Laboratory for Non-equilibrium Synthesis and Modulation of Condensed Matter, Shanxi Province Key Laboratory of Advanced Functional Materials and Mesoscopic Physics, School of Physics, Xi'an Jiaotong University, 710049, Xi'an, Shaanxi, P.R.China}
\author{Xiangru Tao}
\affiliation{MOE Key Laboratory for Non-equilibrium Synthesis and Modulation of Condensed Matter, Shanxi Province Key Laboratory of Advanced Functional Materials and Mesoscopic Physics, School of Physics, Xi'an Jiaotong University, 710049, Xi'an, Shaanxi, P.R.China}
\author{Yundi Quan}
\email{yundi.quan@gmail.com}
\affiliation{MOE Key Laboratory for Non-equilibrium Synthesis and Modulation of Condensed Matter, Shanxi Province Key Laboratory of Advanced Functional Materials and Mesoscopic Physics, School of Physics, Xi'an Jiaotong University, 710049, Xi'an, Shaanxi, P.R.China}
\author{Peng Zhang}
\email{zpantz@mail.xjtu.edu.cn\\}
\affiliation{MOE Key Laboratory for Non-equilibrium Synthesis and Modulation of Condensed Matter, Shanxi Province Key Laboratory of Advanced Functional Materials and Mesoscopic Physics, School of Physics, Xi'an Jiaotong University, 710049, Xi'an, Shaanxi, P.R.China}

\begin{abstract}
The intermetallic compound TlBi$_2$ crystallizes 
in the MgB$_2$ structure and becomes superconducting 
below 6.2 K. Considering that both Tl and Bi have 
heavy atomic masses, it is puzzling why TlBi$_2$ 
is a conventional phonon-mediated superconductor.
We have performed comprehensive first-principles 
calculations of the electronic structures, the 
phonon dispersions and the electron-phonon couplings 
for TlBi$_2$. 
The $6p$ orbitals of bismuth dominate over the 
states near the Fermi level, forming strong intra-
layer $p_{x/y}$  and inter-layer $p_z$ $\sigma$ 
bonds which is known to have strong electron-phonon
coupling. In addition, the large spin-orbit 
coupling interaction in TlBi$_2$ increases its
electron-phonon coupling constant significantly. 
As a result, TlBi$_2$, with a logarithmic phonon
frequency average one tenth that of 
MgB$_2$, is a phonon-mediated superconductor.
\end{abstract}
\maketitle

\section{Introduction}
The discovery of phonon-mediated superconductivity in MgB$_2$ with a transition temperature of 39 K was an important breakthrough in the search for high-temperature superconductivity.\cite{MgB2}  Theoretical calculations based on density functional theory suggest that boron $p$ states are dominant near the Fermi level and they form two types of bonds, the in-plane $\sigma$ bonds between nearest neighbor (n.n.) $p_{x/y}$ orbitals and the out-of-plane $\pi$ bonds between n.n. $p_z$ orbitals.\cite{PhysRevLett.86.4366,PhysRevLett.87.087005, mgb2choinature} The in-plane $\sigma$ bonds are strongly coupled to lattice vibrations, resulting in large electron-phonon coupling ($\lambda$ $\sim$ 1) and the high superconducting transition temperature (T$_c=39$ K). In contrast, the coupling between the $p_z$-$\pi$ bonds and high frequency boron vibrations is relatively weak. The disparity in terms of coupling strength between $\sigma$ and $\pi$ bonds leads to the well-known two-gap superconductivity in MgB$_2$. \cite{souma2003, mgb2choinature} 

Since the discovery of MgB$_2$, there have been efforts to 
design MgB$_2$-like superconductors by searching for light-element compounds with $p$ states near the Fermi level.
The rationale is that having light elements helps
increase phonon frequencies, while having the in-plane
$\sigma$-bonds among $p$ orbitals near the Fermi level enhances
electron-phonon coupling strength.
Over the years, many MgB$_2$-like compounds have been
discovered and they
can be broadly divided into two groups:
compounds that crystallize in the MgB$_2$ structure and compounds 
that have $\sigma$ states near the Fermi level. 
Examples in the former group include 
vanadium doped HfB$_2$/ZrB$_2$ (8.33/7.31 K)\cite{PhysRevB.95.094505}, 
and silicon doped YbGa$_2$ (2.4 K) \cite{YbGaSi}. A recent high-throughput study by Yu {\it et al} 
systematically analyzed the structural stability and the electron-phonon coupling
of many binary compounds that crystallize in the MgB$_2$ structure under ambient pressure
and several candidate MgB$_2$-like superconductors were proposed.\cite{PhysRevB.105.214517}
For the latter group,
one prominent example among many others is the hole-doped diamond (T$_c$ $\sim$ 4 $K$) 
\cite{dopeddiamond, PhysRevLett.93.237003, PhysRevLett.93.237002} which becomes superconducting after the carbon $\sigma$ bands becomes partially occupied via hole doping. Finding superconductors by searching for the metallization of $\sigma$ bands has
also inspired the recent study on La$_3$Ni$_2$O$_7$ (80 K under pressure).\cite{sun2023superconductivity}

So far, efforts to design MgB$_2$-like superconductors are  centered around light-element compounds, while compounds that
consist of heavy elements are usually considered unfavorable for phonon-mediated superconductivity. However, a recent experiment by Yang {\it et al}\cite{Yang.PhysRevB.106.224501} found that TlBi$_2$ which crystallizes in the MgB$_2$ structure becomes superconducting below 6.2 K. Given that both Tl and Bi have relatively large atomic masses, superconductivity in TlBi$_2$ is unexpected. Understanding the origin of the superconductivity in TlBi$_2$ is therefore important for future theoretical and experimental searches for MgB$_2$-like superconductors. 

As atomic number increases, the spin-orbit coupling (SOC) strength increases as well. The atomic numbers of thallium and bismuth are 81 and 83 respectively, which are fairly large. Therefore, the spin-orbit coupling interaction is expected to have important impacts on the electronic structures, the electron-phonon coupling constants and the topological properties of TlBi$_2$. Lead (Pb), which sits between thallium and bismuth in the periodic table, is known to have SOC-enhanced electron-phonon coupling constant. \cite{PONCE2016116}. Whether SOC has a similar impact on TlBi$_2$ is still unknown and we will explore it in this paper. In addition, SOC is also known to lift band degeneracies and to create topological band composites that could possibly host non-trivial topological states. These topological states are especially important in superconductors because of their potential to host Majorana Fermions. \cite{PhysRevB.92.115119} 

In this paper, we present a comprehensive
first-principles study of TlBi$_2$,
in hope of uncovering the origin of its strong electron-phonon coupling.
In addition, the topological properties of TlBi$_2$ are calculated and we point out possible non-trivial topological states in this conventional phonon-mediated superconductor.

\section{Crystal structure and Method}
\begin{figure}[!ht]
    \includegraphics{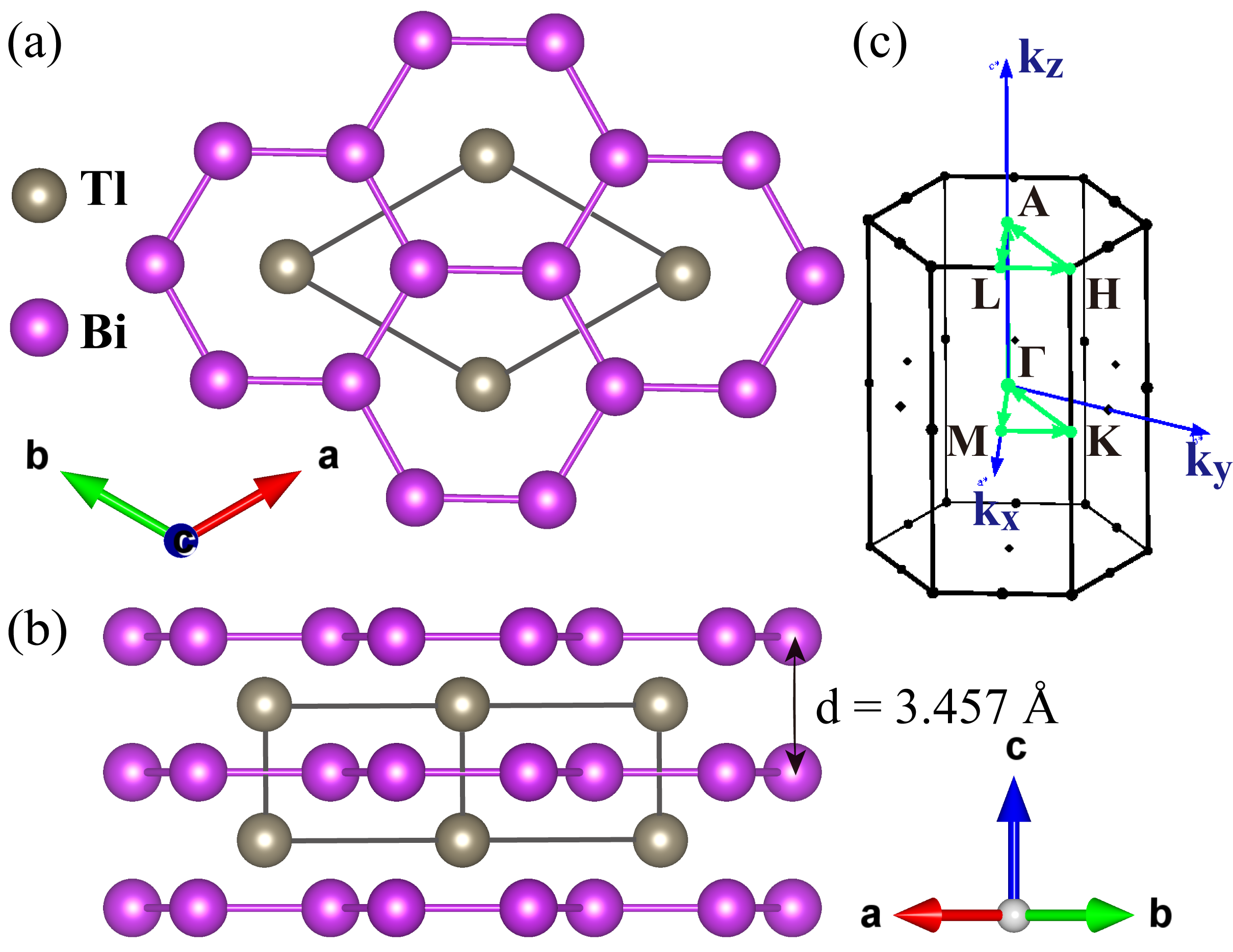}
  \caption{The (a) top- and (b) side-view of the crystal structure of TlBi$_2$. (c) Brillouin zone of TlBi$_2$. Bi atoms form two-dimensional honeycomb layers interlaced with Tl atoms.} 
  \label{cs}
\end{figure}

The intermetallic compound TlBi$_2$ crystallizes in the AlB$_2$-type
structure with Tl interlaced between two-dimensional Bi 
honeycomb layers, as shown in Fig.~\ref{cs} (a) and (b). Its space group is P6/\textit{mmm} (No. 191) and the calculated lattice constants are a = 5.749 \AA and c = 3.457 \AA, which is consistent with experimental parameters\cite{Yang.PhysRevB.106.224501}. The Tl and Bi atoms occupy 1\textit{a} and 2\textit{d} Wyckoff positions, respectively, namely, Tl: (0, 0, 0), Bi: (1/3, 2/3, 1/2). This crystal structure is known to host high temperature superconductivity, e.g. MgB$_2$ (39 K). However, unlike MgB$_2$ which has weak inter-layer interaction, bismuth atoms in TlBi$_2$ form strong inter-layer bonding due to the relatively small $c/a$ ratio.

Density functional theory calculations are carried out using the planewave code Quantum Espresso.\cite{Giannozzi2009,Giannozzi_2017} The exchange-correlation potential is approximated using the generalized gradient approximation (GGA) as parameterized by Perdew, Burke, and Ernzerhof.\cite{PBE} We use the optimized norm-conversing pseudopotential proposed by Hamann.\cite{PhysRevB.88.085117,VANSETTEN201839} The kinetic energy cutoff and the charge density cutoff of the plane wave basis are chosen to be 60 and 240 Ry, respectively. Self-consistent calculations are carried out using a $\Gamma$-centered mesh with $24\times24\times24$ k points and a Methfessel-Paxton smearing width of 0.02 Ry. Structural optimization is performed as well, with 
the convergence threshold on total energy at $1.0\times 10^{-7}$ Ry and that on forces at $1.0\times 10^{-5}$ Ry/Bohr. \\

Phonon dispersions, electron-phonon couplings (EPC), and superconducting transition temperatures are calculated using the density functional perturbation theory (DFPT) with a $3\times3\times3$ mesh of q-points. Given that both Tl and Bi have large atomic numbers, the spin-orbit coupling (SOC) effect is important. Therefore, calculations both with and without SOC are
carried out to help understand the impact of SOC. Wannier functions of Bi-6$p$ and Tl-6$p$ are constructed to obtain an effective tight-binding Hamiltonian and to study the low energy physics of TlBi$_2$. 
 
Superconducting transition temperatures
of conventional phonon-mediated superconductors
can be estimated by using the Allen-Dynes equation\cite{PhysRevB.12.905},
\begin{eqnarray}
T_c = \frac{f_1f_2\expval{\omega_{log}}}{1.20} \exp \left ( -\frac{1.04(1+\lambda)}{\lambda - \mu^\ast - 0.62 \lambda \mu^\ast}\right )
\end{eqnarray}
where $f_1$ and $f_2$ are the strong-coupling correction and the shape correction respectively.\cite{PhysRevB.12.905}. 
\begin{eqnarray}
f_1 & = &\left [ 1 + (\lambda/\Lambda_1)^{3/2}\right ] ^{1/3}\\
f_2 & = & 1+ \frac{(\omega_2/\omega_{log} - 1)\lambda^2}{\lambda^2+\Lambda_2^2}\\
\Lambda_1 & = & 2.46 (1+3.8\mu^{\ast})\\
\Lambda_2 & = & 1.82(1+6.3\mu^\ast)(\omega_2/\omega_{log})
\end{eqnarray}
The electron-phonon coupling constant $\lambda$ is
\begin{eqnarray}
\lambda  & = & 2 \int \frac{\alpha^2F(\omega)}{\omega} d\omega
\end{eqnarray}
And $\alpha^2F(\omega)$ is the Eliashberg function which is defined as
\begin{eqnarray}
\alpha^2F(\omega) = \frac{1}{N_F}\sum |g_{mn}^{\nu}|^2 \delta(\epsilon_{mk})\delta(\epsilon_{n,k+q})\delta(\omega - \omega_{\nu,q})
\end{eqnarray}

\section{Results}
\subsection{Electronic structure}

\begin{figure*}[!ht]
  \includegraphics{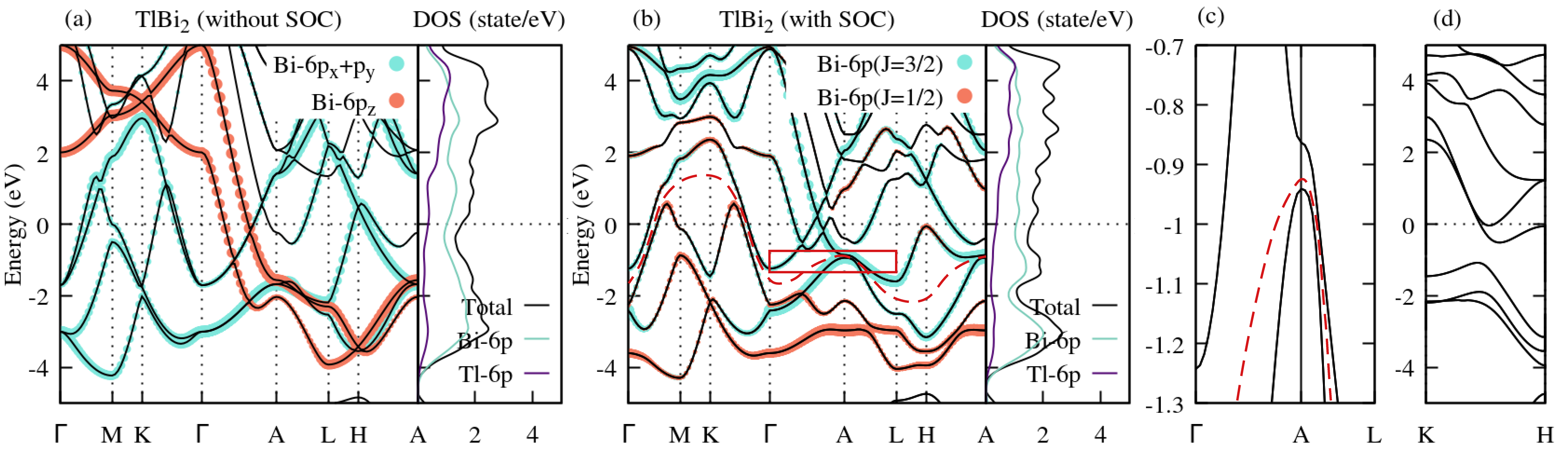}
  \caption{The electronic band structures and the density of states of TlBi$_2$ without (a) and with (b) SOC. (c) Zoom-in view of the solid red box area in (b). (d) The electronic band structures of TlBi$_2$ with SOC along K-H path.}
  \label{es}
\end{figure*}

The band structures and the density of states (DOS) of TlBi$_2$ both with and without SOC are shown in Fig.~\ref{es}. In both cases, the low energy physics of TlBi$_2$ is dominated by the Bi 6$p$ states with a small amount of contribution
from the Tl 6$p$ states. 
In the absence of SOC, two bands (four if spin degeneracy is considered), mostly of $p_{x/y}$ characters, become degenerate near -1.8 eV at the $\Gamma$ point.\cite{PhysRevLett.86.4366} When the SOC effect is included, the degeneracy at $\Gamma$ is lifted resulting in a SOC induced splitting $\Delta_{SOC}^{\Gamma}$ of around 1 eV. Interestingly, the SOC-induced gap opening produces a small gap between band 65 and band 67 in the whole Brillouin zone. In Fig.~\ref{es} (c), the band structure near the $A$ point in the energy range from -1.3 to -0.7 eV is shown to indicate the SOC driven gap opening. 
Although the gap opening due to SOC seems to be small, applying external compressive or tensile strain can effectively increase or decrease the gap, thereby fine tuning the electronic structure of TlBi$_2$.

To identify the topological character of TlBi$_2$, we calculate the $Z_2$ invariants by assuming a "curved chemical potential" (the red dashed line in Fig.~\ref{es} (b) and (c) ) through the gap.\cite{PhysRevB.92.115119} 
The Z$_2$ topological indices for three dimensional
materials with inversion symmetry can be calculated
by inspecting the inversion symmetry operator eigenvalues for the wave functions at the time reversal invariant momentas (TRIMS).\cite{PhysRevB.76.045302} In Table ~\ref{parity}, the parity eigenvalues of TlBi$_2$ for bands \textcolor{red}{61} to \textcolor{red}{66}, along with the Z$_2$ invariants, are given.  
The $Z_2$ invariant of TlBi$_2$ using the curved
chemical potential is (1,000), which suggests
that topologically protected states
could exist between band \textcolor{red}{65} and band \textcolor{red}{67}.  \\

Two studies by Jin {\it et al} \cite{MgB2nl} and Zhou {\it et al} \cite{PhysRevB.100.184511} reported that the electronic structure of MgB$_2$ has symmetry protected
nodal lines along $H \rightarrow K$ when the effect of spin-orbit coupling is neglected. The electronic structure of TlBi$_2$ without considering SOC appears to have degenerate states around -2 eV at the K point and 0.2 eV at the H point that could potentially correspond to the nodal line states. However, due to the large spin-orbit coupling interaction in both thallium and bismuth atoms, these degeneracies are lifted, see Fig.~\ref{es} (d) and a full gap opens up between band 65 and band 67 at every k-point in the Brillouin zone.

\begin{table}[!ht]
    \caption{Parity eigenvalues of band 61, 63 and 65 at time reversal invariant points and the product of parity eigenvalues for all the occupied states.}
    \begin{tabular}{ccccccccccc}
    \hline\hline
    TRIM & Parity & $\delta$ && TRIM & Parity & $\delta$ & \\
    \hline
    $\Gamma$ (0,0,0) & + + $-$ & $-1$ && A (0,0,1/2) & $-$ + $-$ & +1 & \\
    M$_1$ (1/2,0,0) & $-$ + $-$ & +1 && L$_1$ (1/2,0,1/2) & $-$ + $-$ & +1 & \\
    M$_2$ (0,1/2,0) & $-$ + $-$ & +1 && L$_2$ (0,1/2,1/2) & $-$ + $-$ & +1 & \\
    M$_3$ (1/2,1/2,0) & $-$ + $-$ & +1 && L$_3$ (1/2,1/2,1/2) & $-$ + $-$ & +1 & \\
    \hline\hline
    \end{tabular}
    \label{parity}
\end{table}

\begin{figure}[!ht]
  \includegraphics{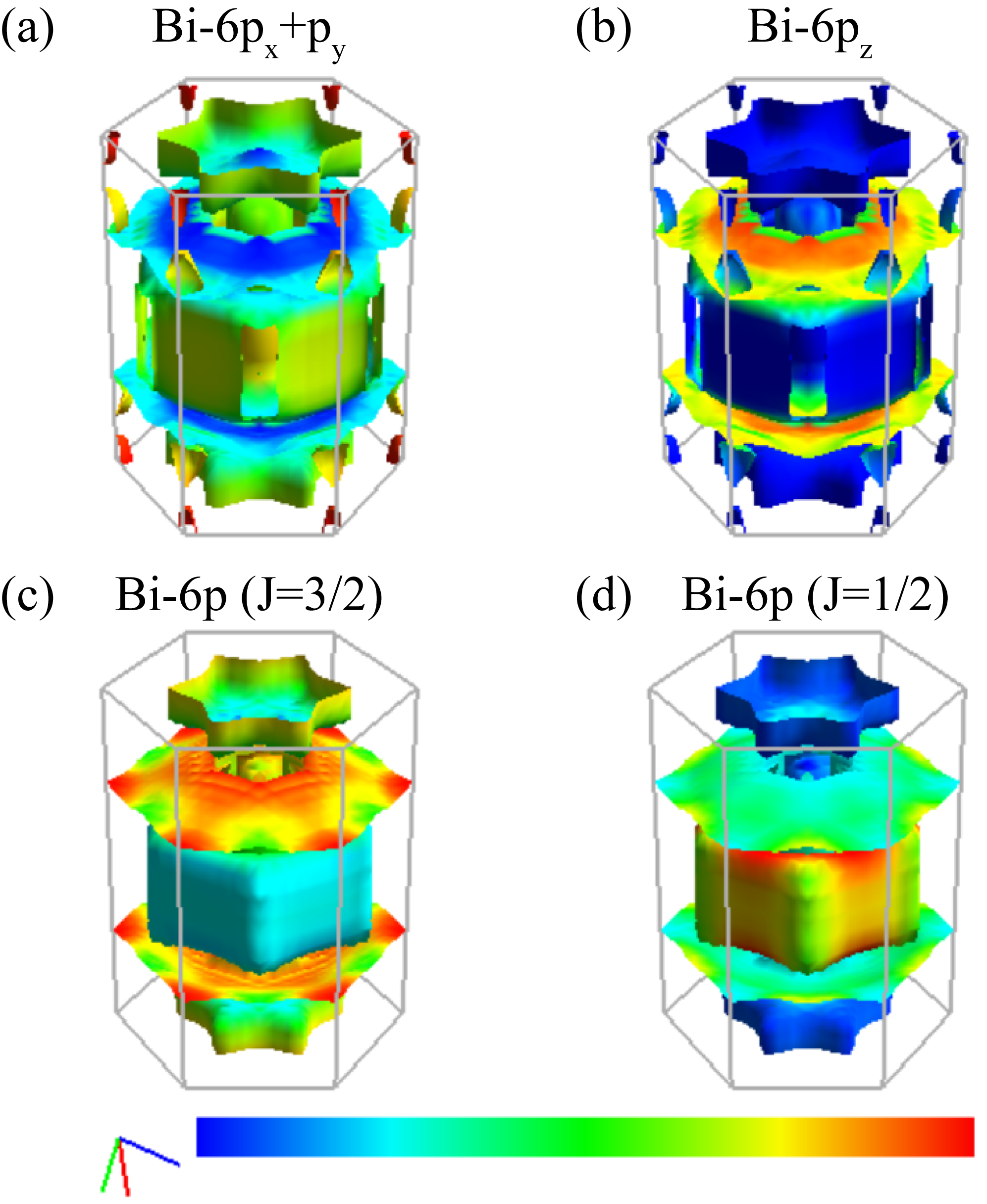}
  \caption{The projections into the nonrelativistic $p_{x,y}$ and $p_z$ basis of the Fermi surface of TlBi$_2$ (calculated without SOC), are presented in panels (a,b), while relativistic orbital character of the Fermi Surface calculated with SOC are presented in panels (c,d). The intensity of projected orbital is illustrated by colorbar, from low (blue) to high (red) values.}
  \label{fs}
\end{figure}

\subsection{Fermi surface nesting }
The importance of Fermi surface nesting in understanding high-temperature superconductivity was discussed extensively in the past. \cite{PhysRevB.29.1243,PhysRevB.77.165135,doi:10.1063/5.0081081} It is often noted that large sections of flat
regions of Fermi surfaces indicate sharp
peaks in the nesting function and consequently large electron-phonon coupling. The Fermi surfaces of TlBi$_2$ that are calculated with and without SOC are shown in Fig.~\ref{fs}. Although the spin-orbit coupling has significant impact on the band structure of TlBi$_2$, it nonetheless leaves the Fermi surfaces almost intact. 

\begin{figure}[!ht]
    \includegraphics[width=\columnwidth]{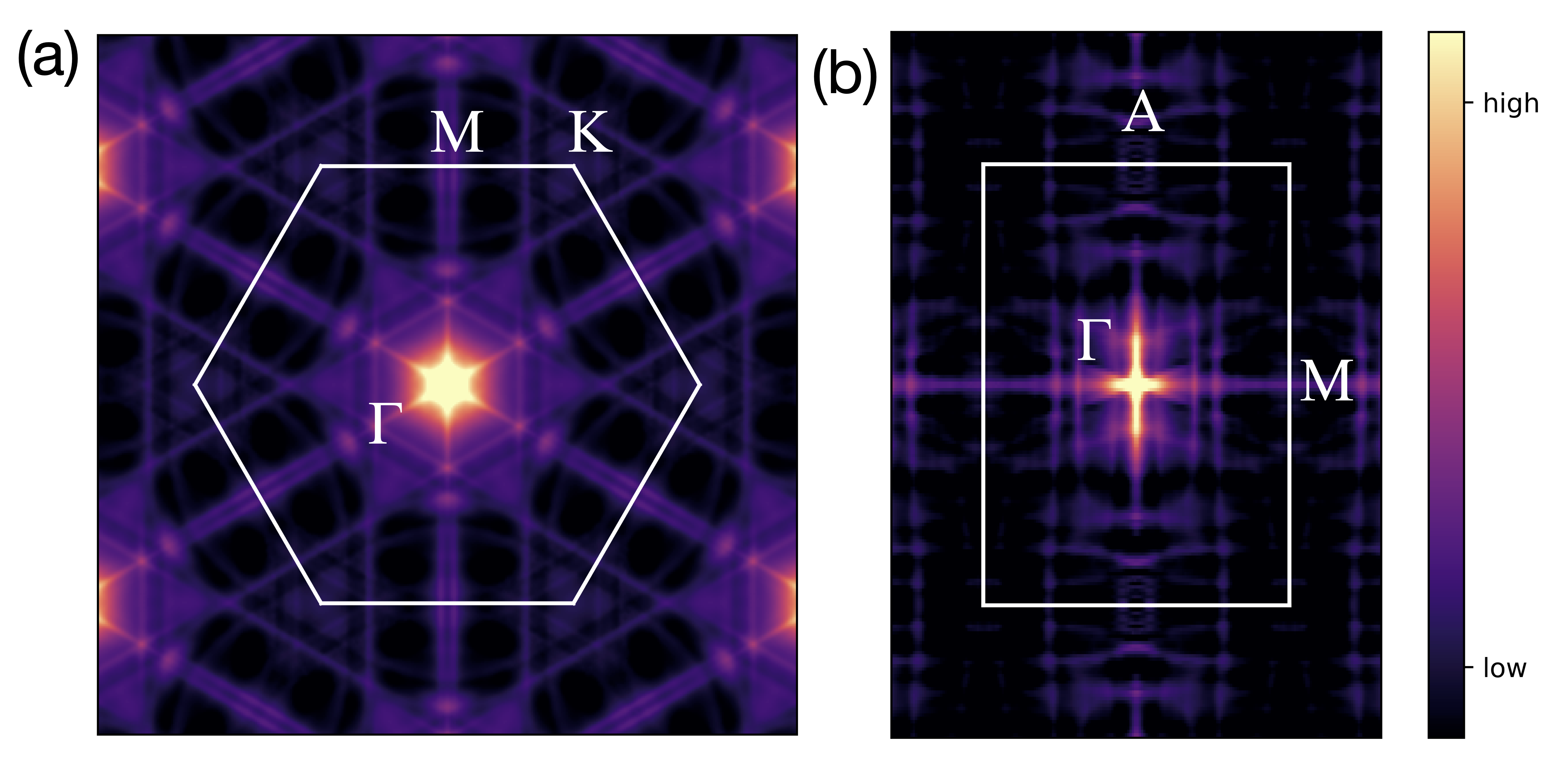}
    \caption{(a) and (b) are the Fermi surface nesting function of TlBi$_2$ on the $k_c=0$ and $k_a=0$ planes respectively. The $k_c=0$, i.e. in-plane nesting, has two peaks between $\Gamma$ and $M$, which originates from the electron-pocket centered at $\Gamma$. The nesting function on $k_a=0$ plane has multiple stripe-like features which can usually be found in two-dimensional materials. } 
    \label{nes}
\end{figure}

The Fermi surfaces shown in Fig.~\ref{fs} are colored by the corresponding orbit characters, from low (blue) to high (red). There is a cage-like electron pocket at the center of the Brillouin zone and its main character is Bi $J=\frac{1}{2}$ states ($p_{x/y}$ if SOC is not included). Right above and below the electron pocket, there are two relatively flat pieces of Fermi surfaces that are dominated by Bi $J=\frac{3}{2}$ states, ($p_z$ if SOC is not included). Near the $A$ point, the Fermi surface is also mostly of Bi $J=\frac{3}{2}$ character, ($p_{x/y}$ if SOC is not included). Overall, the shapes and the sizes of the Fermi surfaces are complex and it might have multiple wave vectors that can give rise to strong
nesting. We therefore calculate the Fermi surface nesting function of
TlBi$_2$ using the EPW code on a 100$\times$100$\times$100 k-mesh. 
In Fig.~\ref{nes}, the nesting functions of TlBi$_2$ on the  $k_c=0$ plane and the $k_a=0$ plane are plotted. Near the middle point between $\Gamma$ and $M$ on the $k_c=0$ plane, there are two nesting function peaks which might originate from the electron Fermi surface pocket at the $\Gamma$ point. There are multiple stripes of nesting function peaks on the $k_a=0$ plane, notably at $k_c=0.3$ and $0.4$. Their origin are likely due to the flat Fermi surfaces above and below the electron pocket at $\Gamma$ and also the Fermi surfaces near the $A$ point.

\subsection{Hopping parameters}

\begin{figure}[!ht]
  \includegraphics[]{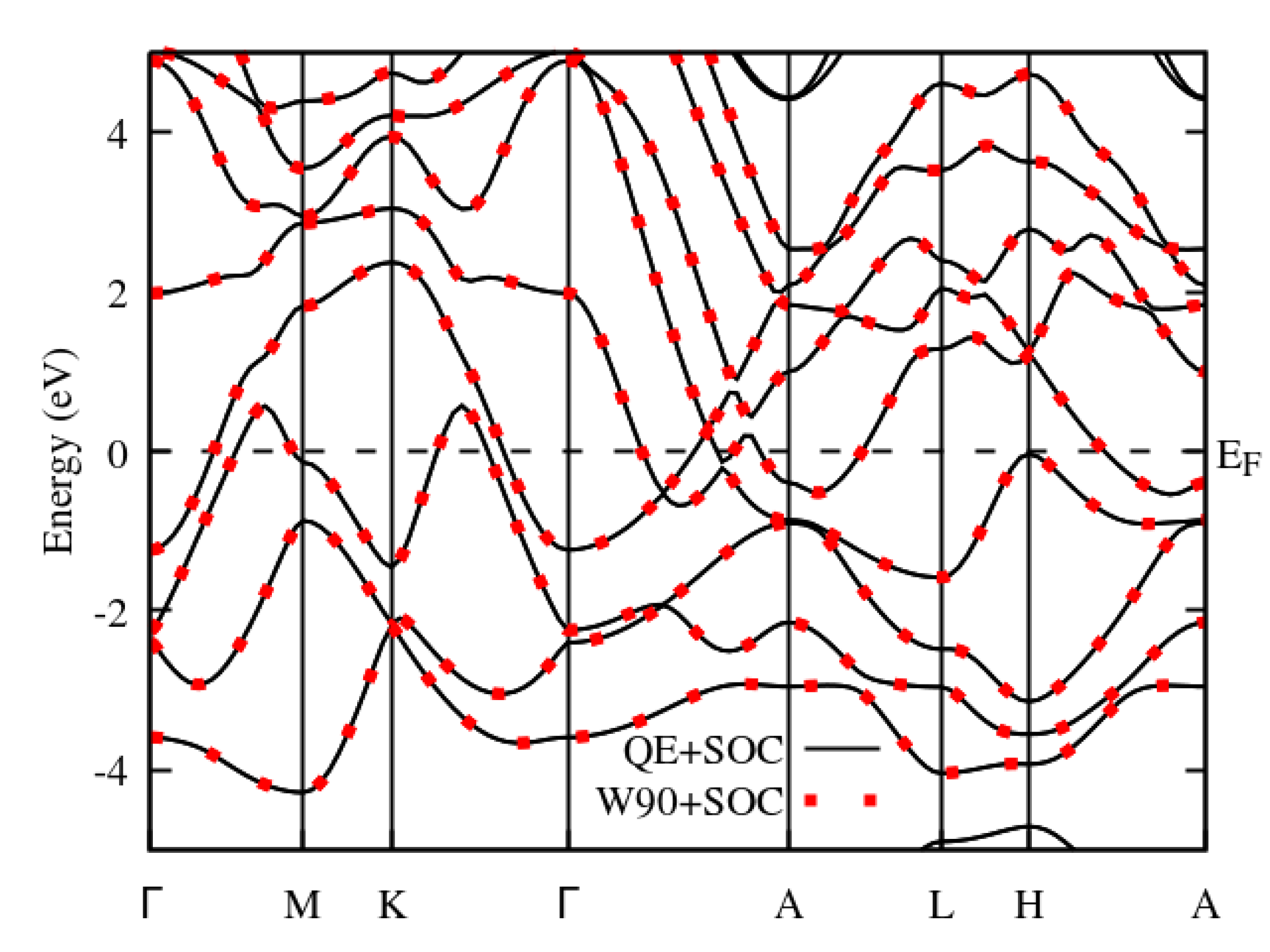}
  \caption{Bandstructure of TlBi$_2$ with SOC. Interpolation with Wannier90 (red dots) and DFT reference bandstructure (solid black)}.
  \label{w90bands}
\end{figure}

\begin{figure}[!ht]
  \includegraphics[]{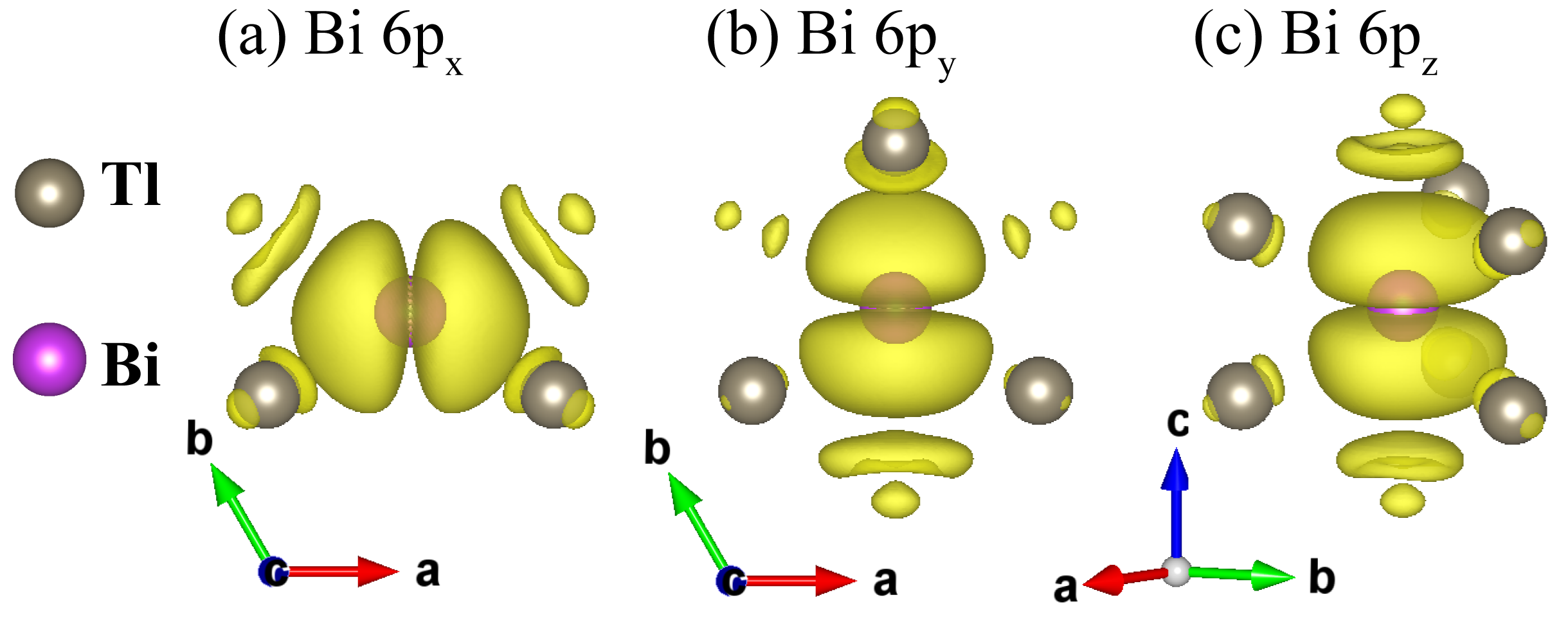}
  \caption{MLWFs in TlBi$_2$ describing the Bi p orbitals with SOC.}
  \label{MLWFs}
\end{figure}

\begin{table}[!ht]
 \begin{tabular}{c|cc|cccccccc}
    \hline
 &&& \multicolumn{2}{c}{Hopping parameter (eV) }\\
 \cline{4-5}
 & \multicolumn{2}{c|}{Wannier orbitals} & TlBi$_2$ & MgB$_2$ \\
 \cline{4-5}
 \hline
 \hline
\multirow{4}{*}{On-site} & X-$p_z$ & X-$p_z$ & 2.79 & 10.71 \\
& X-$p_{x,y}$ & X-$p_{x,y}$  & 1.92 & 8.72  \\
 & M-$p_z$ & M-$p_z$         & 0.37 & 3.57  \\
 & M-$p_{x,y}$ & M-$p_{x,y}$ & 0.36 & 5.62  \\
 \hline
\multirow{3}{*}{Intra-layer} 
& M-$p_x$ &  M-$p_x$ & 1.15 & 2.79  \\
& M-$p_x$ &  M-$p_y$ & -0.70 & -2.91  \\
& M-$p_y$ &  M-$p_y$ &  0.34 & -0.53   \\
\hline
\multirow{4}{*}{X-M}& X-$p_z$ & M-$p_z$  & 0.30 & 0.83  \\
& X-$p_z$ & M-$p_y$ & 0.35 & 1.39  \\
& X-$p_y$ & M-$p_z$ & 0.52 & 1.54   \\
& X-$p_y$ & M-$p_y$ & 0.87 & 0.75   \\
\hline
\multirow{2}{*}{Inter-layer}  & X-$p_z$ & X-$p_z$ & \textbf{1.70} & 0.05 \\
& M-$p_z$ & M-$p_z$  &\textbf{1.37} & -0.27  \\
 \hline
  \end{tabular}
\caption{The hopping parameters among X-$p$ and M-$p$ Wannier orbitals of TlBi$_2$ and MgB$_2$, where X = Tl, Mg and M = Bi, B.}
  \label{hopping}
\end{table}

To gain a microscopic understanding of the electronic structures of TlBi$_2$, we have obtained the hopping
parameters of the states near the Fermi level by calculating the maximally localized Wannier
functions (MLWFs) using the Wannier90 code.\cite{wannier90} The initial projections are chosen to be the Tl and Bi $p$ orbitals. To verify that the Wannier functions can faithfully reproduce the DFT electronic structures, we compare the Wannier-interpolated band structures with the DFT band structures in Fig.~\ref{w90bands} which shows that the interpolated bands are consistent with the DFT results. The Wannier functions that are localized at the Bi atoms are shown in Fig.~\ref{MLWFs}. These MLWFs retain the Bi $p$ orbital characters, suggesting that hybridization between Bi $p$ states and its neighboring Tl states is weak.

In Table ~\ref{hopping}, the hopping parameters of TlBi$_2$ with SOC and MgB$_2$ are given. The on-site energies 
of the bismuth $p_{x, y, z}$ in TlBi$_2$ are around 0.36 eV, while the on-site energies of the boron $p_z$ and $p_{x/y}$ states in MgB$_2$ are 3.57 and 5.62 eV. Compared with MgB$_2$, the intra-layer $\sigma$ bonding among Bi $p_{x/y}$ orbitals is slightly weaker. But the inter-layer $\sigma$ hopping between the $p_z$ orbitals of the Bi atoms sitting at different layers is as large as 1.37 eV, which is in sharp contrast to the negligible inter-layer hopping among boron $p_z$ orbitals in MgB$_2$.

\subsection{Phonon dispersion and electron-phonon coupling}
\begin{figure}[!ht]
  \includegraphics[]{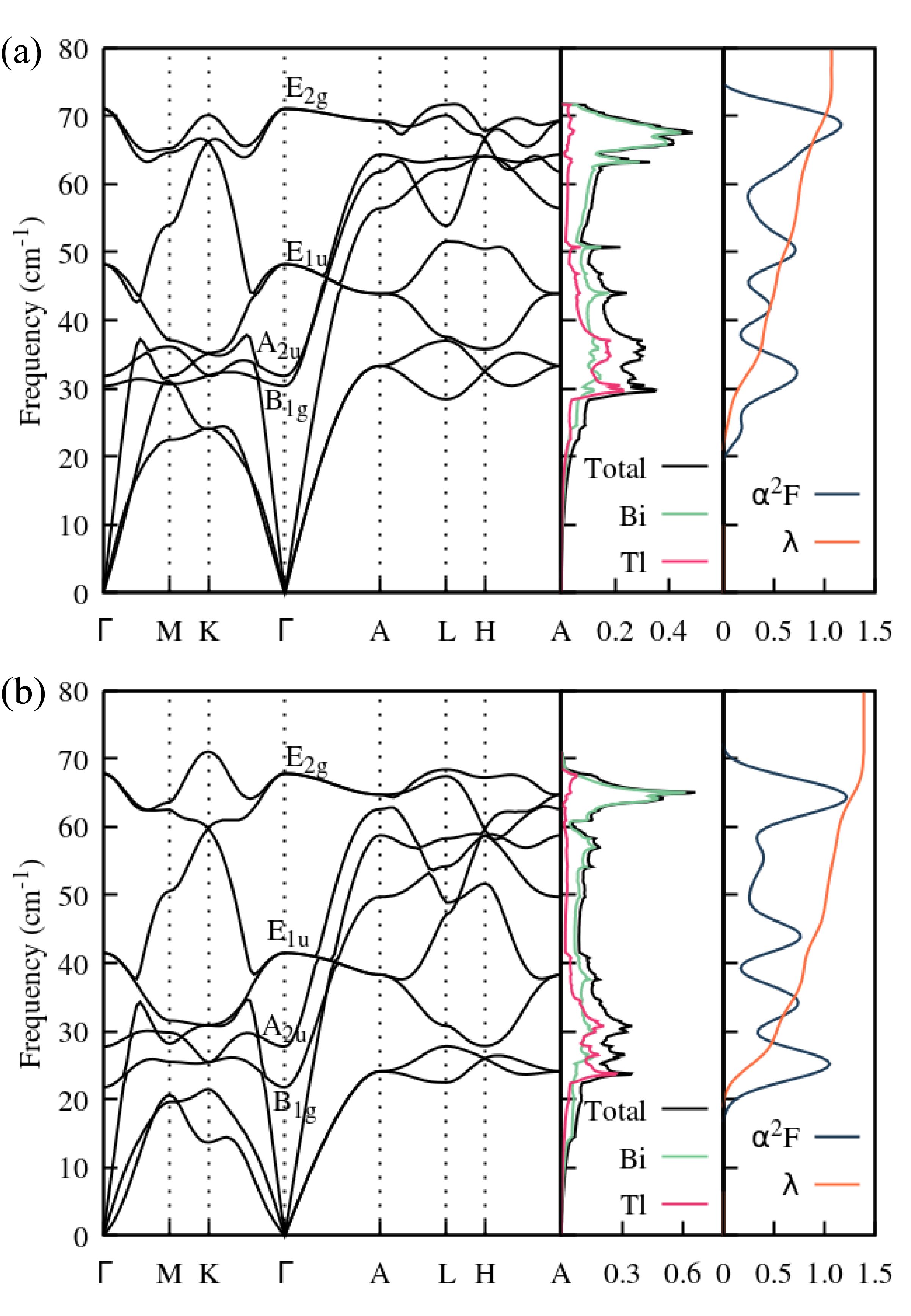}
  \caption{The phonon dispersion relation, the total and projected phonon density of states (PHDOS) and the Eliashberg function $\alpha^2F(\omega)$ of TlBi$_2$ without (a) and with (b) SOC are shown.}
  \label{ph}
\end{figure}
By using the modified McMillan formula, Yang {\it et al}
estimated the electron-phonon coupling constant of TlBi$_2$ 
to be around 1.38.\cite{Yang.PhysRevB.106.224501} In this paper, we calculate the 
electron-phonon coupling constant of TlBi$_2$ by 
running linear-response calculations using the 
Quantum Espresso code. Two calculations, one without SOC 
and the other with SOC, are 
carried out. The phonon dispersions, 
the phonon density of states and the Eliashberg function
$\alpha^2F(\omega)$ without and with SOC 
are shown in Fig.~\ref{ph} (a) and (b) respectively. The Tl phonon partial DOS
has the majority of its weight in the 30 to 40 cm$^{-1}$  range, while Bi vibrations dominate the phonon DOS from 40 to 70 cm$^{-1}$. Therefore, Tl and Bi atoms are active in different phonon frequency ranges.

At $\Gamma$ point, the irreducible representations of the 
six optical modes are $B_{1g}$ (1), $A_{2u}$ (1), $E_{1u}$ 
(2) and $E_{2g}$ (2). The integers inside the braces 
indicate the degeneracy of the corresponding phonon mode. 
In the absence of SOC, the vibrational frequencies of the
$A_{2u}$, $B_{1g}$, $E_{1u}$ and $E_{2g}$ modes are around 30, 32,
48 and 71 $cm^{-1}$
respectively. 
When SOC is
included, the vibrational frequencies of $B_{1g}$, $A_{2u}$,
$E_{1u}$ and $E_{2g}$ at the $\Gamma$ point become 22, 28, 42
and 68 ($cm^{-1}$), which indicates that the spin-orbit
interaction has the effect of phonon softening in TlBi$_2$. As a result, the logarithmic average of the phonon frequencies of TlBi$_2$ decreases from 44 ($cm^{-1}$) to 37 ($cm^{-1}$).
The electron-phonon coupling constant, on the other hand, increases from 1.1 (without SOC) to 1.4 (with SOC). Using the Allen-Dynes equation and the $\mu^\ast$ ranging from 0.1 to 0.13, we obtain T$_c$=4.4-5.1K (without SOC) and 5.0-5.5 K (with SOC).

\section{Discussion and summary}
\begin{table}[!ht]
    \begin{tabular}{cccccccccc}
    \hline\hline
         & \multicolumn{2}{c}{TlBi$_2$} & MgB$_2$  \\
         & w/o SOC & w/ SOC &  & &\\
    \hline
    \hline
$a$ (\AA) & \multicolumn{2}{c}{5.749} & 3.086 &  \\
$c/a$ & \multicolumn{2}{c}{0.601} & 1.142 & \\
$d_{in}$ (\AA) & \multicolumn{2}{c}{3.319} & 1.78  \\
$d_{out}$ (\AA) & \multicolumn{2}{c}{3.457} & 3.52  \\
$d_{out}/d_{in}$ & \multicolumn{2}{c}{1.04} & 1.98  \\
$N(E_F)$ ($eV^{-1}$) & 1.5 & 1.6 & 0.74 &  \\
$\lambda$ & 1.1 & 1.4 & 0.87 &  \\
$\omega_{log}$ ($cm^{-1}$) & 44 & 37 & 504  & \\
$T_c$ (K) & 4.4-5.1 & 5.0-5.5 & 40 &  \\
Ref. & \multicolumn{2}{c}{This paper} & Ref. 29\cite{PhysRevB.64.020501} \\
\hline
  \end{tabular}
    \caption{The structure parameters, the intra-layer and inter-layer distance of Bi/B atoms, the DOS at the Fermi level $N(E_F)$, $\lambda$, $\omega_{log}$ and T$_c$ for TlBi$_2$ and MgB$_2$.}
  \label{comparison}
\end{table}
In comparison with MgB$_2$, the logarithmic average of phonon
frequencies of TlBi$_2$ is low, about an order of magnitude lower than MgB$_2$, see Table~\ref{comparison}. However, its electron-phonon coupling is as large as 1.4 which helps improve its T$_c$. Several factors are at play which helps enhance the electron-phonon coupling strength of TlBi$_2$ and consequently its T$_c$. 1). the inter-layer Bi-Bi $\sigma$ bonding among Bi $p_z$ orbitals is strong and it gives rise to large Fermi surface nesting at a few $q_z$ planes. 2). although bismuth is heavier than thallium, the phonon modes in the frequency region above 40 cm$^{-1}$ are dominated Bi vibrations. And the phonon modes in this region contribute about one third of the total electron-phonon coupling strength. 3) the linear-response calculations without and with soc indicates that SOC can significantly increase the electron-phonon coupling constant $\lambda$ from 1.1 to 1.4. 

Another interesting aspect of TlBi$_2$ is that TlBi$_2$ is a strong 
topological "insulator" using a curved-chemical-potential approach. The relatively large spin-orbit coupling interaction opens up a small gap
between band 65 and band 67 over the whole Brillouin zone. And the size of the
gap can be controlled by applying external strain. Given that there is growing interest in studying the topological properties of MgB$_2$ and similar compounds in 
recent years,\cite{PhysRevLett.123.077001,PhysRevB.101.161407,PhysRevB.100.184511,MgB2nl,PhysRevB.100.094516} TlBi$_2$ could be an interesting candidate for future studies.

{\it Acknowledgment:} This work is supported by National Natural Science Foundation of China No. 11604255 and the Natural Science Basic Research Program of Shaanxi No. 2021JM-001. The computational resources are provided by the HPC center of Xi’an Jiaotong University.


\end{document}